\documentclass[12pt]{amsart}
\usepackage[mathscr]{eucal}
\usepackage{amsmath,amsfonts,amssymb,amsthm}
\usepackage{hyperref}
\usepackage[matrix,arrow]{xy}
\usepackage{times}
\advance\textheight\topskip
\renewcommand{\tilde}{\widetilde}

\newcommand{\bref}[1]{\textbf{\ref{#1}}}

\newcommand{\p}[1]{|#1|}
\newcommand{\gh}[1]{\mathrm{gh}(#1)}
\newcommand{\agh}[1]{\mathrm{antigh}(#1)}
\newcommand{\pgh}[1]{\mathrm{puregh}(#1)}

\newcommand{\cl}{\mathrm{cl}}

\newcommand{\map}{\,\mathrm{:}\,}

\newcommand{\dd}{\partial}

\renewcommand{\geq}{\,{\geqslant}\,}
\renewcommand{\leq}{\,{\leqslant}\,}

\newcommand{\binner}[2]{%
  {\langle}\kern-4.15pt{\langle}#1{,}\,#2{\rangle}\kern-4.15pt{\rangle}}
\newcommand{\commut}[2]{[#1{,}\,#2]}
\newcommand{\pb}[2]{\left\{{}#1{},{}#2{}\right\}}
\newcommand{\ab}[2]{\big(#1,#2\big)}

\newcommand{\half}{\mathchoice{%
    \ffrac{1}{2}}{\frac{1}{2}}{\frac{1}{2}}{\frac{1}{2}}}

\newcommand{\ffrac}[2]{\raisebox{.5pt}%
  {\footnotesize$\displaystyle\frac{#1}{#2}$}\kern1pt}

\newcommand{\brst}{\mathsf{\Omega}}

\newcommand{\dl}[1]{\mathchoice{\ffrac{\dd}{\dd #1}}{\frac{\dd}{\dd
      #1}}{\ffrac{\dd}{\dd #1}}{\ffrac{\dd}{\dd #1}}}

\newcommand{\ddl}[2]{\ffrac{\dd #2}{\dd #1}}

\newcommand{\vdl}[1]{\ffrac{{\delta}}{\delta #1}}

\newcommand{\CC}{\mathcal{C}}

\newcommand{\derham}{\boldsymbol{d}}

\newcommand{\manifold}[1]{\mathscr{#1}}
\newcommand{\manX}{\manifold{X}}

\newcommand{\manM}{\manifold{M}}

\def\cC{\mathcal{C}}

\def\cF{\mathcal{F}}

\def\cH{\mathcal{H}}

\def\cP{\mathcal{P}}

\def\@secnumfont{\bfseries}
\def\subsubsection{\@startsection{subsubsection}{3}%
  \z@{.5\linespacing\@plus.7\linespacing}{-.5em}%
  {\normalfont\bfseries}}
\def\paragraph{\@startsection{paragraph}{4}%
  \z@\z@{-\fontdimen2\font}%
  \normalfont\bfseries}
\def\subparagraph{\@startsection{subparagraph}{5}%
  \z@\z@{-\fontdimen2\font}%
  \normalfont\bfseries}

\makeatother

\begin{document}
\pagestyle{myheadings}
\markboth{\textsc{\small Barnich, Grigoriev}}{%
  \textsc{\small BRST extension of the non linear unfolded formalism}}
\addtolength{\headsep}{4pt}

\begin{flushright}\small
ULB-TH/05-06\\[-2pt]
FIAN-TD/05-08\\
\texttt{hep-th/0504119}\\[-3pt]
\end{flushright}

\begin{centering}
  
  \vspace{1cm}  

  \textbf{\Large{BRST Extension of the Non-Linear Unfolded Formalism}}

\vspace{1cm}

\textit{Proceedings of the conference ``Rencontres Mathematiques de
  Glanon'', July 2004, Glanon, France and the school ``Quantum Field
  Theory, Supersymmetry and Higher Spin Fields'', March 2005, Tomsk,
  Russia.}
  
  \vspace{1.5cm}
  
  {\large Glenn Barnich$^{a,*}$ and Maxim Grigoriev,$^{b}$ }

\vspace{1.5cm}

  \begin{minipage}{.9\textwidth}\small \it \begin{center}
    $^a$Physique Th\'eorique et Math\'ematique, Universit\'e Libre de
    Bruxelles and International Solvay Institutes, Campus
    Plaine C.P. 231, B-1050 Bruxelles, Belgium \end{center}
\end{minipage}
    
    \vspace{.5cm}

\begin{minipage}{.9\textwidth}\small \it \begin{center}
    $^b$Tamm Theory Department, Lebedev Physics
    Institute, Leninsky prospect 53, 119991 Moscow, Russia\end{center}
\end{minipage}
    
\end{centering}

\vspace{1cm}

\begin{center}
  \begin{minipage}{.9\textwidth}
    \textsc{Abstract}. We review the construction of gauge field
    theories from BRST first-quantized systems and its relation to the
    unfolded formalism. In particular, the BRST extension of the
    non-linear unfolded formalism is discussed in some details.
  \end{minipage}
\end{center}

\vfill

\noindent
\mbox{}
\raisebox{-3\baselineskip}{%
  \parbox{\textwidth}{\mbox{}\hrulefill\\[-4pt]}}
{\scriptsize$^*$ Senior Research Associate of the National
  Fund for Scientific Research (Belgium).}

\thispagestyle{empty}
\newpage


\section{Introduction}
This contribution is based on the
papers~\cite{Barnich:2003wj,Barnich:2004cr,Barnich:2005ga} and the
talks given by M.G. at the conference ``Rencontres Math\'ematiques de
Glanon'', July 2004, Glanon, France and the school ``Quantum Field
Theory, Supersymmetry and Higher Spin Fields'', March 2005, Tomsk,
Russia.

Hamiltonian BFV-BRST quantization
\cite{Fradkin:1975cq,Batalin:1977pb,Fradkin:1978xi} as a tool for
constructing gauge invariant field theories was originally used in the
context of open string field theory in the mid eighties (see
\cite{Thorn:1989hm} for an early review). Soon thereafter, this
approach was also used to describe higher spin gauge theories at the
free level \cite{Ouvry:1986dv,Bengtsson:1986ys,Henneaux:1987cp}.
Despite several attempts to constructing consistent interactions in
this approach \cite{Cappiello:1989cd,Bengtsson:1988jt}, the full
interacting theory was eventually constructed in the so-called
unfolded formalism
\cite{Vasiliev:1990en,Vasiliev:1992av,Vasiliev:2003ev}.

In the latter approach, the theory is formulated at the level of
equations of motion while constructing a Lagrangian is a separate
problem that usually requires introducing additional structures.
Recently~\cite{Barnich:2004cr}, the relation between the unfolded and
the BRST approach has been understood at the free level: an extended
BRST parent system has been explicitly constructed which gives rise
upon different reductions both to the standard Fronsdal formulation
and to the unfolded form of the equations of motion. From the
first-quantized point of view, the construction of the parent theory
corresponds to a version of extension used in Fedosov
quantization~\cite{Fedosov:1994,Fedosov:1996fu}. More precisely, this version is 
adapted to the quantization of cotangent bundles~\cite{Bordemann:1997er}.

In this contribution, we develop further the general considerations of
\cite{Barnich:2004cr} concerning the interacting case by describing in
more details the BRST extension of the general non-linear unfolded
equations. We explicitly show that this BRST extension is well suited
for the problem of incorporating additional constraints and the
analysis of various reductions.  We also propose a geometrical
interpretation of the BRST extended unfolded formalism in terms of
supermanifolds and show how it generalizes the so-called AKSZ
construction~\cite{Alexandrov:1997kv}.

\section{Generalities on the BRST formalism}\label{sec:BRST}
In this section, we review the BRST construction in the
non-Lagrangian/non-Ha\-mil\-to\-ni\-an case. We want to emphasize here
that the standard construction of the BRST differential (see
e.g.~\cite{HT-book}) does in general not depend on the existence of an
even/odd bracket structure together with a generator (BRST
charge/master action) and can be constructed in terms of constraints
and gauge generators alone. In the exposition we closely
follow~\cite{HT-book}.  

The relevance of the non-Lagrangian generalization of BRST theory was
emphasized recently in~\cite{Barnich:2004cr}, where a non-Lagrangian
version of generalized auxiliary fields~\cite{Henneaux:1990ua} was
developed in order to discuss possibly non-Lagrangian theories which
become Lagrangian after elimination/addition of unphysical degrees of
freedom. The rationale behind this is the idea that no physical
principle can force one to require unphysical dynamics to follow from
a variational principle. This approach has proved useful both in the
general setting and in the context of higher spin gauge theories.
More recently, the BRST theory for non-Lagrangian/non-Hamiltonian
systems was also studied in~\cite{Lyakhovich:2004xd}, where the
non-Hamiltonian BRST formulation was extended to incorporate weak
Poisson structures and to describe their quantization.

\medskip

Consider a phase/configuration space $\manM_0$ and assume the physical
system to be restricted to a submanifold $\Sigma_0 \subset \manM_0$
described by some constraints/equations of motion. For a gauge system,
$\Sigma_0$ is in addition foliated by integral submanifolds (leaves $=$
gauge orbits) of an integrable distribution and all points of a single
integral submanifold describe the same physical state.

In the simplified, finite-dimensional setting we discuss now, let
$\Sigma$ be specified by regular equations $T_a=0$ and the
distribution determined by a set of vector fields $R_\alpha$ on
$\manM_0$, which restrict to $\Sigma$. The integrability condition on
$\Sigma$ takes the form
\begin{equation}\label{eq:R-prop}
  \commut{R_\alpha}{R_\beta}=U^\gamma_{\alpha\beta}R_\gamma+\ldots
\end{equation}
where $\ldots$ denote terms vanishing on $\Sigma$. 
Note that there is some freedom for the
choices of  $T$ and $R$.

The physical degrees of freedom are coordinates on the
reduced space which is the quotient of $\Sigma$ modulo the gauge
orbits. The basic idea of BRST theory (either
Batalin--Fradkin--Vilkovisky in the Hamiltonian or Batalin--Vilkovisky
in the Lagrangian context
\cite{Batalin:1981jr,Batalin:1983wj,Batalin:1983jr,Batalin:1984ss,%
  Batalin:1985qj} ) is to describe physical quantities
as the cohomology of an appropriately constructed BRST differential
instead of explicitly solving constraints and taking the quotient with
respect to the gauge orbits. The construction of the BRST differential
involves two steps.

The first step consists in reducing to $\Sigma$. For simplicity,
constraints $T_a$ are assumed to have at most first order reducibility
relations.  This means that there exist functions $Z_a^A$ such that
$T_a Z^a_A=0$ and matrix $Z^a_A$ has maximal rank on $\Sigma$, so that
that there are no further reducibility relations. We also assume that
$\manM_0$ is not a supermanifold and therefore the constraints and gauge
generators are Grassmann even. The cohomological description of the
reduction is given by the Koszul-Tate complex: one introduces
Grassmann odd variables $\cP_a,\,\p{\cP_a}=1$ and Grassmann even variables
$\rho_A,\,\p{\rho_A}=0$ and considers then the algebra $\cF^0$ of smooth functions on $\manM_0$ with
values in polynomials in $\cP$ and $\rho$.  This algebra is graded
according to the ``antighost number'', $\cF^0=\oplus_{i\geq 0} \cF^0_i$, with
\begin{equation}
  \agh{\cP}=1\,,\qquad \agh{\rho}=2\,.
\end{equation}
The Koszul-Tate differential $\delta\map \cF^0_i\to\cF^0_{i-1}$ is given by
\begin{equation}
  \delta=T_a\dl{\cP_a}+\cP_a Z^a_A\dl{\rho_A}\,.
\end{equation}
Its nilpotency $\delta^2=\delta\delta=0$ follows from the reducibility
identity $T_a Z^a_A=0$. The cohomology of $\delta$ is concentrated in
degree zero and is given by functions on $\Sigma$:
\begin{equation}
  H_0(\delta,\cF^0)=\CC^\infty(\Sigma)\,,\qquad \quad H_i(\delta,\cF^0)=0\quad i> 0\,.
\end{equation}

The next step is the factorization by the gauge orbits. Consider the
space $\cF_0$ of functions on $\manM_0$ with values in polynomials in
some Grassmann odd variables $\cC^\alpha$.  This algebra is graded
according to the ``pure ghost number'', $\cF_0=\oplus_{i \geq 0}
\cF^i_0$, with $\pgh{\cC}=1$. Because of~\eqref{eq:R-prop}, the vector field
\begin{equation}
  \gamma=\cC^\alpha R_\alpha-\half \cC^\alpha\cC^\beta 
U^\gamma_{\alpha\beta}\dl{\cC^\gamma},
\end{equation}
is nilpotent on $\Sigma$, i.e., $\gamma^2$
is proportional to constraints $T_a$. In pure ghost number zero, the cohomology 
of $\gamma$ in
$\CC^\infty(\Sigma)\otimes\bigwedge (C)$ is given by functions on
$\Sigma$ annihilated by $R_\alpha$ (i.e., functions on $\Sigma$ that are
constant along the gauge orbits). Note that in contrast to $\delta$,
higher cohomology groups of $\gamma$ do not vanish in general.

Let us consider now $\cF=\oplus_{i,j\geq 0}\cF^i_j$, the space of
functions on $\manM_0$ with values in polynomials in 
$\cC,\cP,\rho$. Algebra $\cF$ is to be identified with the
algebra of functions on a supermanifold $\manM$
called extended phase (or configuration) space.
The natural degree in $\cF$ is given by the difference
of the pure ghost number and the antighost number: for a
homogeneous element $f$
\begin{equation}
  \gh{f}=\pgh{f}-\agh{f}\,.
\end{equation}
The fact that $\gamma$ is nilpotent up
to terms vanishing on $\Sigma$ can be reformulated as nilpotency of $\gamma$
in the cohomology of $\delta$.
Both vector fields $\gamma$ and $\delta$ can be extended to $\cF$ in
such a way that $\delta$ remains nilpotent while $\gamma$ commutes
with $\delta$. The existence of 
additional vector fields $s_1,s_2,\ldots$ of antighost
number $1,2,\ldots$ such that
\begin{equation}
  s=\delta+\gamma+s_1+s_2+\ldots\,,\qquad \gh{s}=1\,,
\end{equation}
satisfies $s^2=0$ is then guaranteed because the cohomology of
$\delta$ is concentrated in antighost number zero. 

Let us recall how the BRST construction specializes to the case where
$\manM_0$ is the field space of the Lagrangian system described by the
gauge invariant action $S(\phi)$. In this case constraints $T_a$
are equations of motion $\dl{\phi^a}S=0$ while the gage
generators are symmetries of the action i.e., Noether
identities $R_\alpha S=0$ hold. At the same time the Noether identities
imply that equations of motion are not independent: $R^a_\alpha \dl{\phi^a}S=0$.

Because the equations of motion can be considered as components of a
$1$-form on the field space while reducibility identities and the
gauge symmetries are determined by the same generators
$R_\alpha=R_\alpha^a\dl{\phi^a}$ there is a natural odd Poisson
bracket (antibracket) in $\cF$ which is determined by
\begin{equation}
  \ab{\phi^a}{\cP_b}=\delta^a_b\,,\qquad 
  \ab{\cC^\alpha}{\rho_\beta}=\delta^\alpha_\beta\,.
\end{equation}
In this context variables $\cP_a$ and $\rho_\alpha$ are called
antifields and usually denoted by $\phi^*_a,\cC^*_\alpha$. 
Note that the bracket is Grassmann odd and carries ghost
number $1$.  The BRST differential of the Lagrangian gauge system can
be taken to be canonically generated in the antibracket. Namely, for
such an $s$ there exists a generating function $\mathbf{S}\in \cF$ with
$\p{\mathbf{S}}=\gh{\mathbf{S}}=0$ such that
\begin{equation}
  s=\ab{\mathbf{S}}{\cdot\,}\,,\qquad \half\ab{\mathbf{S}}{\mathbf{S}}=0\,.
\end{equation}
The function $\mathbf{S}$ is called master action while the second
equation ensuring nilpotency of $s$ is referred to as the master
equation. The expansion of $\mathbf{S}$ in the antifields reads as
\begin{equation}
  \mathbf{S}=S+\cP_a R^a_\alpha \cC^\alpha+\half\rho_\alpha U^\alpha_{\beta\gamma}\cC^\beta\cC^\gamma+\ldots\,,
\end{equation}
where $\ldots$ denote higher order terms in the expansion with respect to the the antighost number.
This BRST construction and the corresponding quantization method is
known as BV (Ba\-ta\-lin--Vil\-ko\-vis\-ky) formalism.

Finally, let us recall the BRST construction in the case where $\manM_0$
is the phase space of a first-class constrained system.  In this case
$\manM_0$ is equipped with a symplectic structure and the gauge
generators are Hamiltonian vector fields generated by the constraints,
i.e.
\begin{equation}
  R_\alpha=\pb{T_\alpha}{\cdot}\,,
\end{equation}
where $\pb{\cdot}{\cdot}$ is the Poisson bracket determined by the
symplectic structure on $\manM_0$. Because the constraints and the gauge
generators are not any longer independent, the Poisson bracket can be
naturally extended to $\cF$ by defining
\begin{equation}
  \pb{\cP_\alpha}{\cC^\beta}=-\delta^\alpha_\beta\,,
\end{equation}
with the bracket between ghosts and other phase space variables
vanishing. In this case, the BRST differential $s$ is
canonically generated in the extended Poisson bracket: there exists
$\Omega^\cl\in \cF$ such that
\begin{equation}
  s=\pb{\Omega^\cl}{\cdot}\,,\qquad \half\pb{\Omega^\cl}{\Omega^\cl}=0\,.
\end{equation}
The expasion of $\Omega^{\cl}$ in ghost momenta $\cP$ reads as
\begin{equation}
  \Omega^{\cl}=\cC^\alpha T_\alpha+\half\cP_\gamma U^\gamma_{\alpha\beta}\cC^\alpha\cC^\beta+\ldots\,.
\end{equation}
If in addition the Hamiltonian $H^\cl_0\in\cF^0_0$ satisfies $R_\alpha H^\cl_0=T_
\beta V_\alpha^\beta$
for some $V^\alpha_\beta$ (i.e. $H^\cl_0$ is gauge invariant on
$\Sigma$), one can construct 
a BRST invariant Hamiltonian $H^\cl$ such that
\begin{equation}
\pb{\Omega^\cl}{H^\cl}=0\,,\quad \gh{H^\cl}=0\,,\quad
H^\cl|_{\cP=\cC=0}=H^\cl_0\,. 
\end{equation}
This Hamiltonian BRST construction and the corresponding quantization
method is referred to as BFV (Batalin--Fradkin--Vilkovisky) formalism.

\section{Gauge theories associated to the first-quantized BFV-BRST systems}
\subsection{Operator BFV quantization}
\label{BRSTop}

At the quantum level phase space observables become operators represented in a
space of states. The BRST operator $\brst$ and the Hamiltonian operator $H$
satisfy
\begin{equation}
\half\commut{\brst}{\brst}=0,\qquad
\commut{\brst}{H}=0\,,
\end{equation}

Physical operators are described by hermitian operators $A$ 
such that $\commut{\brst}{A}=0$
where two such operators have to be identified if they differ by a
BRST exact operator $A\sim A+\commut{\brst}{B}$.
These two equations define the BRST operator cohomology.
Note that if an inner product is specified
on the space of states, hermitian conjugation is just standard conjugation in this inner product.
In general, however, conjugation can be defined as an additional structure even if
no suitable inner product is given.

Similarly, physical states are selected by the condition $\brst\phi=0$.
Furthermore, BRST exact states should be considered as trivial, or
equivalently, states that differ by BRST exact ones should be
identified $\phi\sim\phi+\brst\chi$. These two equations define the BRST state cohomology.
Finally, time evolution is described by the Schr\"odinger equation
\begin{eqnarray}
i \frac{d \phi}{d t}= H \phi\label{4}.
\end{eqnarray}

\subsection{Free gauge theory on the space of quantum states}\label{sec:3.2}

We consider a quantum BFV-BRST system which we assume to be
time-reparametriza\-tion invariant so that its Hamiltonian $H$ vanishes. 
For a detailed discussion of the general case we
refer to~\cite{Barnich:2003wj}.

We also assume that among the degrees of freedom, there are coordinates
$x^\mu$ which are interpreted as coordinates of a space-time
manifold $\manX_0$ and which are quantized in the coordinate representation. The
space of states is then given by functions of $x^\mu$ taking values
in an internal space $\cH$. (In a geometrically nontrivial situation
one should consider sections of a suitable vector bundle over space-time instead.) 
Because we are not interested in constructing proper quantum mechanics, we are not concerned
with normalizability of the states. For simplicity, we thus can consider states with a smooth
dependence on $x^\mu$.

Given a graded superspace $\cH$, one associates to each basis vector
$e_A$ a coordinate $\psi^A$ with ghost number $\gh{\psi^A}=-\gh{e_A}$
and Grassmann parity $\p{\psi^A}=\p{e_A}$. One then considers
$\manM_\cH$ to be the supermanifold with coordinates $\psi^A$. In what
follows we denote by $\psi^{A_k}$ the fields associated with the ghost
number $-k$ subspace $\cH^{(k)}\subset \cH$, which implies in
particular that $\gh{\psi^{A_k}}=k$. We also introduce the object
$\Psi(x)=e_A \psi^A(x)$ with $\p{\Psi}=0,\gh{\Psi}=0$, called string
field, which is in particular useful to avoid using indices
(see~\cite{Barnich:2004cr} and references therein for precise
  definitions and the relation with the similar
  notion~\cite{Gaberdiel:1997ia} used in the context of string field
  theory).

The configuration space of the free field theory associated with the
quantum system is the space of maps from $\manX_0$ to the submanifold
$\manM_\cH^{(0)}$ of ghost number zero fields. In terms of coordinates
it is described by fields $\psi^{A_0}(x)$. The equations of motion are
given by
\begin{equation}\label{eq:eom}
  \brst^{A_{-1}}_{B_0} \psi^{B_0}(x)=0\,.
\end{equation}
Due to $\brst^2=0$, these equations are invariant under the gauge transformations
$\delta_\epsilon\psi^{A_0}=\brst^{A_0}_{B_1}\epsilon^{B_1}$,
for some gauge parameters $\epsilon^{B_1}$.

The fields associated with states in nonzero ghost number are to be interpreted
as ghost fields, ghosts for ghosts, and antifields for the BRST-BV description of
the theory. 
The BRST-BV differential $s_0$ on the fields $\psi^A(x)$ is then defined by 
\begin{equation}\label{brstwj}
s_0\Psi=\brst\Psi\,\quad\Longleftrightarrow\quad s_0\psi^A=\brst^A_B\psi^B.
\end{equation}

If in addition we are given with an inner product that makes $\brst$ formally self-adjoint,
one can build a classical action and the Batalin-Vilkovisky master action which
determine the equations of motion and the BRST differential respectively~(for more details, 
see~\cite{Barnich:2004cr} and references therein).

\subsection{Algebraic structure of interactions}\label{subsec:algebra}

We now briefly discuss the general algebraic structures underlying
consistent deformations of a theory described by a linear BRST
differential $s_0$ given in \eqref{brstwj}.  For simplicity, we
use here De Witt's condensed notation consisting in including the
space-time coordinates $x^\mu$ in the index $A$ of the coordinates
$\psi^A$ on $\manM_\cH$.

A general nonlinear deformation $s$ of $s_0=\brst_A^B\psi^A\dl{\psi^B}$ has the form
\begin{equation}
s=\brst_A^B\psi^A\dl{\psi^B}+U_{AB}^C \psi^A\psi^B \dl{\psi^C}+
U_{ABC}^D\psi^A\psi^B \psi^C \dl{\psi^D}+\ldots\,,
\end{equation}
with $\p{s}=1$, $\gh{s}=1$, and $\ldots$ denoting higher
order terms. The deformation is consistent if $s^2=0$.
Such a nilpotent vector field $s$ on a flat supermanifold associated with a superspace
$\cH$ is equivalent to an $L_\infty$ algebra on $\cH$~\cite{Lada:1993wc}.

\bigskip 

The simplest example is provided by an $s$ that is at most quadratic. In this case, $\cH$ is a differential graded Lie algebra with $\commut{e_A}{e_B}=e_C
U_{AB}^C$. In the more general case, the $U^A_{BC}$
determine a Lie algebra structure only in cohomology of $\brst$ and there are higher order brackets related to the higher orders terms in $s$.

In the case of field theories, some complications arise because $\cH$
is an infinite dimensional space of field configurations and the
algebraic structures are really represented by differential operators.
Moreover, space-time locality of the deformation is to be taken into
account.  However, in some cases one can explicitly separate the
space-time dependence in such a way that interactions do not involve
an explicit $x$-dependence or $x$-derivatives. We now turn to the
discussion of systems of this type.

\section{Geometry of the Vasiliev unfolded formalism}

\subsection{First-quantized BRST picture}

Consider the special class of free theories associated with a BRST
first quantized model for which fields are defined on a supermanifold
$\manX$ with Grassmann-even coordinates $x^\mu$ 
and Grassmann-odd coordinates $\theta^\mu$. The latter coordinates are  space-time ghosts with $\gh{\theta^\mu}=1$ and can be identified
with the basic differentials $dx^\mu$. The fields take
values in a supermanifold $\manM_\cH$ associated with the graded
superspace $\cH$ and the coordinates on $\manM_\cH$ are denoted
by $\Psi^A$. The components in the expansion of $\Psi^A(x,\theta)$ in
$\theta^\mu$ can be considered as differential forms on $\manX_0$:
\begin{equation}
  \Psi^A(x,\theta)=(\psi_0)^A(x)+\theta^\mu(\psi_1)^A_{\mu}(x)+
\,\theta^\mu\theta^\nu\,(\psi_2)^A_{\mu\nu}(x)+\ldots\,,
\end{equation}
with $\gh{(\psi_p)^A_{\mu_1\ldots\mu_p}}=\gh{\Psi^A}-p$
and $\p{(\psi_p)^A_{\mu_1\ldots\mu_p}}=\p{\Psi^A}-p\,\,mod~2$.
We also assume that the BRST differential $s_0$ can be represented
in the form
\begin{equation}
  s_0\Psi(x,\theta)= \derham\Psi(x,\theta)+\bar\brst\Psi(x,\theta)
\end{equation}
with $\derham=\theta^\mu\dl{x^\mu}$ and $\bar\brst$ a linear
operator in $\cH$, i.e. $\bar\brst\Psi=e_A\bar\brst^A_B\Psi^B$. Note that
the system described by $s_0$ is explicitly space-time covariant.

Under sufficiently general conditions, one can show that nonlinear
deformations of the theory preserving the general covariance can be
assumed to contain neither $x^\mu$ and $\theta^\mu$ derivatives nor an
explicit dependence on these
variables~\cite{Brandt:1996mh,Barnich:1996mr}. The deformed
differential $s$ is then determined by an odd vector field $Q$ on
$\manM_\cH$
\begin{equation}
Q=\bar\Omega_B^A\Psi^B\dl{\Psi^A}+\Psi^B\Psi^C U_{BC}^A\dl{\Psi^A}
+ \Psi^B \Psi^C \Psi^D U_{BCD}^A\dl{\Psi^A}+\ldots\,,
\end{equation}
with $\gh{Q}=\p{Q}=1$ and satisfying the compatibility condition
$Q^2=0$. In other words, $\cH$ is equipped with an $L_\infty$ algebra structure.

Given such a $Q$, the BRST differential itself is then determined by
\begin{equation}
\label{eq:BRST-nl}
s\Psi^A=\derham \Psi^A+Q^A(\Psi)\,.
\end{equation}
The dynamical equations of the system determined by
$s$ are
\begin{equation}
\label{eq:phys-unfold}
  \left(\derham\Psi^A+Q^A(\Psi)\right)\Big|_{\psi^{(l)}=0,\,l\neq 0}=0\,
\end{equation}
where we have put to zero all the component fields
$(\psi_p)^A_{\mu_1\ldots\mu_p}(x)$ entering $\Psi(x,\theta)$ except those
of ghost number zero.  In the case where $\gh{\Psi^A}\geq 0$, this
is exactly the form of the general unfolded equations proposed
in~\cite{Vasiliev:1988xc,Vasiliev:1988sa,Vasiliev:1994gr}.  Equations
of this form are also known as defining the structure of a free
differential algebra~\cite{Sullivan}. 

Some comments are in order.
Note that for each coordinate function $\Psi^A$ of ghost number
$\gh{\Psi^A}={p_A}$, there is at most one component field
$(\psi_{p})^A_{\mu_1\ldots\mu_p}$ with a given ghost number. Note also
that if $p_B<0$ for some $B$, then equations~\eqref{eq:phys-unfold}
reduce to the constraint equations, $Q^B(\Psi)|_{\psi^{(l)}=0,\,l\neq
  0}=0$ because in $\Psi^B$ there is no ghost number zero component field
so that the first term in~\eqref{eq:phys-unfold} vanishes.

The BRST differential also determines gauge transformations for physical
fields.  Let $e_{a}$ be a basis in the subspace of $\cH$ with
zero or negative ghost number, i.e., $\gh{e_a}\leq 0$ so that the
associated coordinates $\Psi^{a}$ carry nonnegative ghost numbers.  It
then follows that among component fields in the expansion of
$\Psi^{a}$ with respect to $\theta^\mu$ there is a field
$\psi^a_{\mu_1\ldots\mu_p}$ with $p=\gh{\Psi^a}$ of zero ghost number.
The gauge transformation of $\psi^a$ is given by
\begin{equation}
  \delta_\epsilon \psi^{a}=s\psi^{a}\big|_{\psi^{(l)}=0,\, l\neq 0,1}
\end{equation}
with the ghost number $1$ fields $\psi^{(1)}$ replaced by gauge
parameters $\epsilon$. Observing that the right hand side is linear
in $\psi^{(1)}$ and of the same form degree as $\psi^{a}$, one arrives
at a more explicit form for the gauge transformations:
\begin{equation}
  \delta_\epsilon \psi^{a}=\derham \epsilon^a
+\epsilon^A\ddl{\Psi^A}{Q^a}\Big|_{\psi^{(l)}=0,\,l\neq 0}\,.
\end{equation}

The BRST differential~\eqref{eq:BRST-nl} can naturally be considered as an
extension of the unfolded equations~\eqref{eq:phys-unfold}. It 
allows for cohomological tools to be used at the level of the field theory, e.g., for the introduction or elimination of generalized (non-Lagrangian) auxiliary fields~\cite{Henneaux:1990ua,Barnich:2004cr}. 

Arbitrary unfolded equations can be embedded in such a BRST
system. Indeed, suppose that the equations of motion of a system are given by 
\begin{equation}
\label{eq:ueq}
d\psi^{A}+Q^A(\psi)=0\,,
\end{equation}
where $\psi^A$ are differential forms on $\manX_0$ with form degree
denoted by $p_A$. Suppose furthermore that $Q^A$ are
polynomial (in the sense of the wedge product of differential forms)
functions in $\psi^A$ satisfying the compatibility condition
\begin{equation}
\label{eq:comp-unf}
Q^B\frac{\partial Q^A}{\partial \psi^B}=0.
\end{equation}
For simplicity we assume that these compatibility conditions are
satisfied without making use of the relations of the Grassmann algebra
of basis $1$-forms $dx^\mu$ besides the supercommutativity of
$\psi^A$-s with respect to the wedge product.\footnote{Such free
  differential algebras are called \textit{universal}, for details see
  e.g.~\cite{Bekaert:2005vh}.} This can be equivalently formulated in
terms of an auxiliary linear supermanifold $\manM$ with independent
coordinates $\Psi^A$ with
$\p{\Psi^A}=\p{(\psi^{p_A})^A_{\mu_1\ldots\mu_{p_A}}}+p_A$ and $\gh{\Psi^A}=p_A$ as the
nilpotency condition $Q^2=0$ for an odd vector field
\begin{equation}
  Q=Q^A(\Psi)\dl{\Psi^A}\,,
\end{equation}
on $\manM$. One can then introduce additional fields on $\manX_0$ to define 
superfields $\psi^A(x,\theta)$  on $\manX$ taking values in $\manM$ with
\begin{equation}
\gh{(\psi^{p_A})^A_{\mu_1\ldots\mu_{p_A}}}=\gh{\Psi^A}-p_A\,,\qquad 
\p{(\psi^{p_A})^A_{\mu_1\ldots\mu_{p_A}}}=\p{\Psi^A}-p_A\,,
\end{equation}
so that the original fields $\psi^A_{\mu_1\ldots\mu_{p_A}}$ appear as
ghost number zero component fields from $\Psi^A(x,\theta)$.  If one
now considers the BRST differential~\eqref{eq:BRST-nl} determined by
$Q$, it is straightforward to verify that the dynamical equations
\eqref{eq:phys-unfold} coincides with the original unfolded
equations~\eqref{eq:ueq}.

\subsection{Geometric picture --  non-Lagrangian AKSZ procedure}
When reformulated in BRST terms, the unfolded equations allow for a
nice geometrical interpretation. Consider two
supermanifolds: a supermanifold $\manX$ equipped with a degree, 
an odd nilpotent vector field $\derham,\,\mathrm{gh}_\manX(\derham)=1$, and a
volume form $d\mu$ preserved by $\derham$ and a
supermanifold $\manM$ equipped with another degree, an odd nilpotent vector field $Q$,
$\mathrm{gh}_\manM(Q)=1$. As implied by the notation,
the basic example for $\manX$ is the odd tangent bundle $\Pi
T\manX_0$ which has a natural volume form and is equipped with the 
De Rham differential. Note that supermanifolds equipped with an
odd nilpotent vector field are often called $Q$-manifolds~\cite{Schwarz:1992gs}.

Consider then the manifold of maps from $\manX$ to $\manM$ (more
generally, one could of course consider the space of sections of a
bundle over $\manX$ with fibers isomorphic to $\manM$). This space is
naturally equipped with the total degree denoted by $\gh{\cdot}$ and
an odd nilpotent vector field $s$, $\gh s=1$. If $z$ are local
coordinates on $\manX$ (in the case where $\manX=\Pi T\manX_0$
coordinates $z$ split into $x^\mu$ and $\theta^\mu$) and $\Psi^A$ are
coordinates on $\manM$, the expression for $s$ reads
\begin{equation}
\label{eq:s-AKSZ}
  s=\int_{\manX}d\mu (-1)^{\p{d\mu}}\left[
\derham\Psi^A(z)+Q^A(\Psi(z))
\right]
\vdl{\Psi^A(z)}\,.
\end{equation}
Vector field $s$ can be considered as a BRST differential of a field
theory on $\manX$. Indeed, the basic properties $s^2=0$ and $\gh{s}=1$
hold.  In what follows we refer to this system as a quadruple
$(\manX,\derham,\manM,Q)$, where manifolds $\manX$ and $\manM$ are
equipped with the odd nilpotent vector fields $\derham$ and $Q$
respectively. In addition, $\manX$ is equipped with a
$\derham$-invariant volume form and the ghost grading on
$\manX$ and $\manM$ is such that $\mathrm{gh}_\manX{(\derham)}=1$ and
$\mathrm{gh}_\manM{(Q)}=1$.

For the system $(\manX,\derham,\manM,Q)$ it is easy to check using the explicit
form~\eqref{eq:s-AKSZ} that $s\Psi^A=\derham\Psi^A+Q^A$. This shows that, 
locally, \eqref{eq:s-AKSZ} describes the same theory as $s$
defined in~\eqref{eq:BRST-nl} if $\manM=\manM_\cH$,
  $\manX=\Pi T\manX_0$, and $\derham =\theta^\mu\dl{x^\mu}$. 
  
In the case where the ``target'' manifold $\manM$ is in addition
equipped with a compatible (odd) Poisson bracket $\pb{\cdot}{\cdot}$
and $Q=\pb{S}{\cdot}$ is generated by a ``master action'' $S$
satisfying the classical master equation $\half \pb{S}{S}=0$, one can
construct a field theory master action $\mathbf{S}$ on the space of
maps. This procedure was proposed in~\cite{Alexandrov:1997kv} as an
approach for constructing BV-BRST formulations of topological sigma
models. Further developments can be found
in~\cite{Cattaneo:1999fm,Batalin:2001fh,Batalin:2001fc,Cattaneo:2001ys,Park:2000au,Roytenberg:2002nu,Edgren:2002xg}
and references therein.  A generalization that also includes the Hamiltonian
BRST formulation has been proposed in~\cite{Grigoriev:1999qz} and
covers the case where $\mathbf{S}$ is Grassmann odd and is to
be interpreted as a BRST charge of the BFV-BRST formulation of the theory.

\subsection{Generalized auxiliary fields}\label{sec:gen-aux}

The restriction that the compatibility condition~\eqref{eq:comp-unf}
holds without making use of the Grassmann algebra relations for the
basic differential forms is not really necessary.  Moreover, in
practice it often happens that there are some other constraints on
$\manM$. Nevertheless, it is still possible to bring the system to the
form~\eqref{eq:BRST-nl} by explicitly solving these constraints or by
appropriately extending $\manM$.

To show how constraints on $\manM$ can be incorporated in the BRST
differential, suppose that we are in the setting of the previous subsection
and let also $\Sigma\subset \manM$ be a submanifold in $\manM$ such
that $Q$ restricts to $\Sigma$.  In terms of some constraints $T_a$ determining $\Sigma$, this
means that $QT_a|_\Sigma=0$. The system described in this way is just a
system without constraints but with $\manM$ replaced with $\Sigma$ and
$Q$ replaced by its restriction $Q|_\Sigma$ to $\Sigma$. For this system to be well defined, it is actually enough to require that $Q^2$ be zero in $\manM$ only up to terms vanishing on
$\Sigma$.

For simplicity, let $T_a$ be independent, regular constraints. One
then introduces variables $\cP_a$ with $\gh{\cP_a}=-1$,
$\p{\cP_a}=\p{T_a}+1$ and extends $\manM$ to
$\manM_{{\cP}}=\manM\times\Lambda$ where $\Lambda$ is a linear
supermanifold with coordinates $\cP_a$. Exactly the same arguments as
in Section~\bref{sec:BRST} then show that one can construct
\begin{equation}
 Q_{{\cP}} = T_a \dl{\cP_a}+Q+Q_0+Q_1+Q_2+\ldots
\end{equation}
satisfying $Q^2_{{\cP}}=0$. Here, $Q_i$ denote terms of degree
$i$ in $\cP_a$. 

We claim that the system $(\manX,\derham,\manM_{{\cP}},Q_{{\cP}})$ is
equivalent to the system
$(\manX,\derham,\Sigma,Q|_{\Sigma})$ through elimination of generalized
auxiliary fields (in the non-Lagrangian sense,
see~\cite{Barnich:2004cr}). Indeed, let $s_\cP$ be a BRST differential of
$(\manX,\derham,\manM_{{\cP}},Q_{{\cP}})$ then $\cP_a$ and $s_{{\cP}}\cP_a$ are
independent constraints because $s_{{\cP}}\cP_a=T_a+\ldots$. Moreover,
equations $s_{{\cP}}\cP_a=0$ at $\cP=0$ are equivalent to $T_a=0$,
while $T_a$ can be taken (locally) as independent fields so that one
concludes that $\cP_a,s_{{\cP}}\cP_a$ are generalized auxiliary fields
and can be eliminated. The reduced system is obviously 
$(\manX,\derham,Q|_{\Sigma},\Sigma)$.

Conversely, let $w^a$ be some constraints on $\manM$ such that
$w^a,Qw^a$ are independent constraints determining a surface $\Sigma
\subset \manM$.  The same arguments then show that the system
$(\manX,\derham,\manM,Q)$ can be reduced to the system
$(\manX,\derham,\Sigma,Q|_{\Sigma})$ through the elimination of
generalized auxiliary fields

Consequently, if one works
in BRST terms, one can assume without loss of generality that all constraints on the fields 
are already incorporated in $Q$, which can be useful from various points of view.
In particular, this also shows that it is enough to consider
the case where $\manM$ is a linear supermanifold with all the nontrivial geometry
encoded in $Q$.

\subsection{Elimination of pure gauge variables in space-time}

As explained in general in Subsection~\bref{sec:3.2}, the field theory
differentials $\derham$ and the operator $\bar\Omega$ determining the
linear part of $Q$ can be understood in first quantized terms as a
BRST operator acting in a space of quantum states.  The equivalence
under elimination of generalized auxiliary fields for $Q$, or more
precisely, its linear part, can then be understood as a natural
equivalence of first quantized systems under elimination of pure gauge
degrees of freedom.

Among the first quantized degrees of freedom, variables
$x^\mu,\theta^\mu$ and their conjugate momenta are represented in the
coordinate representations on functions in $x^\mu$ and $\theta^\mu$
and are identified with space-time coordinates, while the other
degrees of freedom are represented in the target space $\cH$. Of
course one could as well represent $\theta^\mu$ in $\cH$.
This makes no difference because the respective representation space
is finite dimensional. That introduction/elimination of pure gauge
variables represented in $\cH$ leads to theories related by
introduction/elimination of generalized auxiliary fields was shown in
details in~\cite{Barnich:2004cr}. One can expect the same for
  pure gauge variables represented on functions on $\manX$. As we are
  going to see, the respective theories are also related by elimination of
  generalized auxiliary fields if one allows for nonlocal
  transformations in the sector of the pure gauge variables.

Consider then a not necessarily linear system determined by
$\manX,\derham,\manM,Q$ and replace $\manX$ with $\manX \times
\manM_{t,\theta}$ and $\derham$ with
$\derham^\prime=\derham+\theta\dl{t}$. Here, $\manM_{t,\theta}$
denotes a linear supermanifold with coordinates $t,\theta$ with
$\p{t}=0,\,\p{\theta}=1$ and $\gh{t}=0,\,\gh{\theta}=1$. Note
  that from a first quantized point of view, for the free part of the system, 
  this corresponds to adding a pair of pure gauge
  variabels $t,\pi_t$ together with their associated ghost variables
  $\theta,\pi_\theta$ with commutation relations
  $\commut{\pi_t}{t}=-1$, $\commut{\pi_\theta}{\theta}=-1$ and adding
  the respective term $\theta\pi_t$ to the BRST charge. These pure gauge
  variables are represented on functions of $t,\theta$ so that the
  additional term in the BRST charge acts as $\theta\dl{t}$.

The resulting system is again a system of the same type, but living on
the extended space--time manifold $ \manX \times \manM_{t,\theta}$,
and the question arises as to how it relates to the original system.
To be able to compare these two field theories, we first need to
consider them as field theories determined on the same space--time
manifold. To this end we identify $(\manX\times
\manM_{t,\theta},\derham+\theta\dl{t},\manM,Q)$ with
$(\manX,\derham,\manM^\prime,Q^\prime)$ where $\manM^\prime$ and
$Q^\prime$ are the configuration space and the BRST differential of
the system $(\manM_{t,\theta},\theta\dl{t},\manM,Q)$. In other words,
the space-time coordinate $t$ becomes a continuous index for fields on
$\manX$, while $\theta\dl{t}$ becomes a part of the target space BRST
differential $Q^\prime$.

The supermanifold $\manM^\prime$ can then be identified with the
manifold of smooth maps from $\manM_{t,\theta}$ to $\manM$ while the
BRST differential $Q^\prime$ is determined in coordinates by
\begin{equation}
  Q^\prime \Psi^A(t,\theta)=\theta\dl{t}\Psi^A(t,\theta)+Q^A(\Psi(t,\theta))\,.
\end{equation}
On $\manM^\prime$, it is useful to take the coordinates
$\tilde\Psi^A,\Psi^A_{0t},\Psi^A_{1t}$ with $\Psi^A_{0t}\big|_{t=0}=0$ so that
a general map has the form
\begin{equation}
  \Psi^A(t,\theta)=\tilde\Psi^A+\Psi^A_{0t}+\theta\Psi^A_{1t}\,.
\end{equation}
It then follows that fields $\Psi^A_{0t}(z)$ and $\Psi^A_{1t}(z)$ on
$\manX$, with  $z$ denoting coordinates on $\manX$, are generalized
auxiliary fields.  Indeed, in terms of the coordinates
  $\tilde\Psi^A,\Psi^A_{0t},\Psi^A_{1t}$, the term in $Q^\prime$
corresponding to $\theta\dl{t}$ acts as
$\int\,dt(\dl{t}\Psi^A_{0t})\vdl{\Psi^A_{1t}}$.  This shows that at
$\Psi^A_{1t}=0$, equations $Q^\prime\Psi^A_{1t}=0$ takes the form
$\dl{t}\Psi^A_{0t}=0$ which has as unique solution $\Psi^A_{0t}=0$
taking into account $\Psi^A_{0t}\big|_{t=0}=0$.  The arguments from the end of
the subsection~\bref{sec:gen-aux} then show that the fields
$\Psi^A_{0t}(z)$ and $\Psi_{1t}(z)$ are indeed generalized auxiliary fields
and can be eliminated, showing the equivalence to the original system
on $\manX$. For systems in unfolded form, the possibility to
add/eliminate space time coordinates together with their differentials
was first observed in the context of higher spin gauge theories
in~\cite{Vasiliev:2001zy} (see
also~\cite{Vasiliev:2003ar,Gelfond:2003vh,Engquist:2002gy} for a more
recent discussion).

More generally, if on $\manX$ one can find constraints $t^\alpha$ such
that $t^\alpha$ and $\derham t^\alpha$ are independent, similar
arguments show that, locally in space-time, one can consistently reduce
the theory to the ``constraint surface'' $\tilde\manX\subset \manX$
determined by the constraints $t^\alpha,\derham t^\alpha$. In the case
where $\derham=\theta^\mu\dl{x^\mu}$, this means that locally in $\manX$
one can consistently reduce the theory to a point. From the BRST
theory point of view, this can also be understood as a version of the
statement that, for theories of this type, representatives of various
cohomology groups can be taken not to depend on space-time
derivatives~\cite{Brandt:1996mh,Barnich:1996mr}.  Hence, cohomology
groups are described by functions, tensor fields, etc. on the target
space $\manM$.  In particular, this implies that possible consistent
deformations and conserved currents of the system are determined by
appropriate $Q$-cohomology classes in $\manM$.  \bigskip

As a final remark, we comment on the BRST extension of the unfolded
formalism as a generally-covariant first-order formalism.
Indeed, manifold $\manM$ can be considered as 
the space of initial data for the equations of motion, while the equations
determine a multi-parametric flow, the number of parameters being the space-time dimension.
If one mods out by the constraints and the gauge freedom, this
multi-parametric evolution is uniquely determined by the initial data. 

As an illustration, it is useful to consider the one-dimensional case
which corresponds to a time reparametrization-invariant Hamiltonian
system. Such a system is described by a BRST charge $\Omega$ and a
symplectic structure on the phase space $\manM$ with coordinates
$\Psi^A$.  The BV-BRST extension of the dynamics is governed by the
master action
$\mathbf{S}$~\cite{Batalin88,Siegel:1989nh,Batlle:1989if,Fisch:1989rm}
\begin{equation}
  \mathbf{S}=\int\, dt d\theta(V_A(\Psi)\theta\dl{t}\Psi^A-\Omega(\Psi))\,,
\end{equation}
which we wrote in the superfield form proposed
in~\cite{Grigoriev:1999qz}. Here $V_A$ is the symplectic potential and
$\theta$ is the superpartner of the ``time'' variable $t$ with
$\p{\theta}=1,\gh{\theta}=1$ (for details and precise definitions
see~\cite{Grigoriev:1999qz,Barnich:2003wj}).  The BRST differential
determined by $\mathbf{S}$ can be written as
\begin{equation}
  s\Psi^A=\derham\Psi^A+Q^A(\Psi)\,,\qquad
 \derham=\theta\dl{t}\,,\quad Q^A=\pb{\Omega}{\Psi^A}\,,
\end{equation}
where $\pb{\cdot}{\cdot}$ is the Poisson bracket corresponding to the
symplectic structure on $\manM$. On the one hand, $s$ is just the standard
BRST differential of the BV formulation for the reparametrization-invariant Hamiltonian system on
$\manM$ written in terms of superfields.  On the other hand, it
can be considered as the BRST differential describing the one-dimensional system
in unfolded form. This illustrates, in particular, the role of space-time coordinates
in the unfolded formalism.  They play exactly the same role as an
evolution parameter in the Hamiltonian formulation of
time-reparametrization invariant systems.

\bigskip

\noindent
\textit{Note added:} While this contribution was being completed, there
appeared reference~\cite{Vasiliev:2005zu} where, among other things,
related aspects of the unfolded formalism are also discussed.

\subsection*{Acknowledgments}
M.G. is grateful to M. Vasiliev for illuminating discussions.  The
work of G.B. is supported in part by a ``P{\^o}le d'Attraction
Interuniversitaire'' (Belgium), by IISN-Belgium, convention 4.4505.86,
by the National Fund for Scientific Research (FNRS Belgium), by
Proyectos FONDECYT 1970151 and 7960001 (Chile) and by the European
Commission program MRTN-CT-2004-005104, in which the author is
associated to V.U.~Brussel. The work of M.G. was supported by the RFBR
Grant 05-01-00996 and by the Grant LSS-1578.2003.2.

\providecommand{\href}[2]{#2}\begingroup\raggedright\endgroup

\end{document}